# Measuring Object-Oriented Design Principles

Johannes Braeuer

Advisor: Reinhold Ploesch
Department of Business Informatics – Software Engineering
Johannes Kepler University Linz
Altenberger Str. 69, 4040 Linz, Austria
johannes.braeuer@jku.at

*Abstract*— The idea of automatizing the assessment of object-oriented design is not new. Different approaches define and apply their own quality models, which are composed of single metrics or combinations thereof, to operationalize software design. However, single metrics are too fine-grained to identify core design flaws and they cannot provide hints for making design improvements. In order to deal with these weaknesses of metric-based models, rules-based approaches have proven successful in the realm of source-code quality. Moreover, for developing a well-designed software system, design principles play a key role, as they define fundamental guidelines and help to avoid pitfalls. Therefore, this thesis will enhance and complete a rule-based quality reference model for operationalizing design principles and will provide a measuring tool that implements these rules. The validation of the design quality model and the measurement tool will be based on various industrial projects. Additionally, quantitative and qualitative surveys will be conducted in order to get validated results on the value of object-oriented design principles for software development.

## I. Introduction

Design principles are fundamental guidelines that help software designers and developers to build and maintain a software system [1] [2]. In contrast to design patterns, whose applicability highly depends on the project or application-domain context, design principles are more generally applicable, supporting designers in building a common consensus about architectural and design knowledge [3]. Furthermore, design principles support beginners in software engineering to avoid traps and pitfalls in object-oriented design [3]. Examples of design principles include the Single Responsibility Principle (SRP), Don't Repeat Yourself Principle (DRY), Separation of Concerns (SOC), and Liskov Substitution Principle (LSP), to name just a few of the 39 design principles we have identified so far.

In addition, design principles help constructively in building sustainable software. Having well-educated architects and designers who understand these principles, enhances the design quality of source code through the proper application of these guidelines. Regardless of the power of design principles in development, we need an objective way to identify deviations from design principles that manifest themselves in the source code for the purposes of measurement and analysis. One general measurement approach relies on single metrics or simple combinations thereof. However, these metrics are too fine-grained to be interpreted in isolation [4]. Metrics can only raise an alert, but they cannot indicate problems in object-oriented software design. Hence, only measuring design quality based on metrics, makes it difficult to identify the core problems and lacks the ability to provide valuable hints for effective and efficient improvements to the design.

The goal of this research is the validation of object-oriented design principles in an industrial context in order to investigate the usefulness and general acceptance of these principles. This research is based on a rule-based quality reference model for operationalizing design principles and a measurement tool. This measurement tool is used to automatically verify the compliance of the source code with the rules and — on a higher level of abstraction — to verify the compliance of the source code with the design principles. Both, the reference model and the measurement tools will be enhanced and completed during this research. The research method of the thesis involves the assessment of the applicability of the design model and toolset in the context of projects from various industrial partners.

## II. Contribution and Research Questions

In [5], the authors point out that organizations involved in object-oriented software development should invest in the definition of design standards. To support this task, the thesis will contribute to the following software-design challenges.

### A. Reference quality model for object-oriented design

With the realization of the value of design assessments and their role in evaluating and improving design quality, there is a need for a design-quality reference model [6]. By identifying this lack of measurement in software design, we are going to use design principles as one major concept for building a reference model; further referred to as design quality model. Design principles are the core of our design assessments because we believe that they are the key for disclosing design problems and we are not aware of any work that tries to systematically measuring them based on a solid validation. Our design quality model will be operational in the sense that the measurement of the design principles is based on a set of rules instead of single metrics as practiced by other approaches. By using rules instead of metrics, we think that a software architect or engineer will gain deeper insight into the non-conformity of a design with the principles and will gain far better guidance for removing flaws and improving object-oriented design in general.

*B. Tool for the automatic measurement of design principles*

Based on the design quality model, a comprehensive software tool will be provided to support the automatic measurement of design principles. In other words, the specified rules of the design quality model will be operationalized by this measuring tool, which helps to identify violations of the design principles. A first version of this measurement tool (unpublished work) is the basis of further development in the context of this thesis.

*C. Acceptance of design principles in software engineering*

This thesis' most central contribution focuses on the usefulness of following design principles during software development. We have already conducted a survey to investigate the importance and relevance of different design principles (the results are not yet published); however, we are not aware of their application or significance in practice. As a result, the work here will answer some open questions regarding the use of design principles in an industrial context.

Based on the above challenges, the thesis will investigate the following research questions:

*RQ1. How can rules help to enhance software design?* We consider it as necessary to understand the value, the importance, and the trustworthiness of each rule. The value of a rule is related to the development support it provides. In contrast, the importance is more focused on the impact of a rule on good object-oriented design. Finally, the trustworthiness of a rule concentrates on the quality of the technical implementation expressed by the false-positive and the true-positive rate.

*RQ2. Are the design principles completely measured by their associated rules?* We will investigate the relation between each design principle and the rules for measuring them. It is therefore necessary to find out whether the design principle is completely covered by its rule set or whether there remain unmeasured or unmeasurable aspects of the design principle. This has to be made explicit and in best-case added (maybe only as manual measures) to the design quality model.

*RQ3. Are design principles crucial for software development?* From a more general viewpoint, we also want to understand the purpose of design principles for the process of designing software architectures and developing software systems. We already know that some design principles are more important than others, but up to now we remain unaware of the value of their application in an industrial software-engineering context.

III. MEASURING DESIGN PRINCIPLES

In order to demonstrate our proposed approach for measuring design principles, we first discuss the design quality model of design principles and then afterwards outline the tool for operationalizing this model.

*A. Design quality model*

A set of 39 design principles has already been identified by a literature study and well specified by us in an internal Wiki. Despite the high number of principles, the previously mentioned survey already indicates that approximately seven out of the 39 principles seem to have higher practical relevance than the others.

The current version of our design quality model (yet unpublished work) is based on the factor model developed in the Quamoco project [7]. This quality model has already proven successful in an industrial context, so we can rely on a fundamental knowledge base [8]. By following this proposed modeling approach, a good, comprehensive reference model can be derived by enhancing the existing model that addresses the design of object-oriented systems.

*B. Implementation of the measuring tool*

Figure 1 provides an overview of our tool architecture for the operationalization of design principles. Understand[1] is a commercial tool that parses source code (e.g., Java source code) and stores the extracted information about types, methods, method parameters, dependencies between classes and packages, and so on in a database. Furthermore, the tool Understand provides a Perl-based API for accessing this stored information.

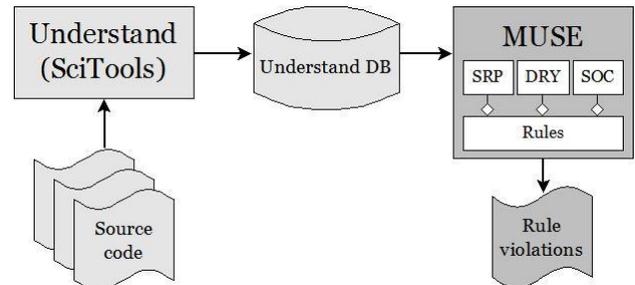

Figure 1: Tool architecture for operationalizing design principles

MUSE (**M**USE **U**nderstand **S**cripting **E**ngine) is a Perl library that we developed in a previous project that is enhanced for the purposes of this thesis. It provides basic functionality for accessing the Understand database, for dealing with threshold values, and for writing result files (in Figure 1 denoted as Rule violations), as well as providing a set of helper functions to more easily access the information provided by Understand (e.g., getting the package of a class, getting the list of all public classes of a project, etc.).

Moreover, MUSE uses the modular concept of Perl to separate the implemented rules from each other. In case a module is stored in a specific directory of MUSE, it can be loaded dynamically, after which the rule defined in this module will be executed. This facilitates the configuration of MUSE (i.e., which rules or design principles to measure) depending on the design-quality requirements of the project. Currently, the rule package consists of a set of 65 rules; this set will reach 80+ rules throughout the thesis project.

As shown in Figure 1, MUSE creates an output file, which lists all rule violations. The number of violations depends on the configuration used, threshold values, and of course on the rules that were selected for execution. In order to visualize the rule violations and to make them analyzable without deep understanding of the MUSE toolkit, the findings file can be uploaded to SonarQube [2]. The drilldown mechanisms in SonarQube allow retracing the rule violations by digging into the analyzed source code and identifying the issues.

---

[1] https://scitools.com/

IV. RESEARCH METHOD AND EVALUATION

For addressing the shortcomings of a missing reference quality model for object-oriented design and for evaluating the implemented measuring tool, we will apply quantitative and qualitative research methods based on case studies.

*A. Quantitative*

Specifically, we will use our measuring tool to iteratively assess projects of industrial partners. This assessment of the source code of a completed iteration will then be handed over to the partner so that architects and engineers can discuss the rule violations that were found in the software design. Subsequently, we will conduct a quantitative survey using pre-defined questionnaires. These questionnaires are designed especially to address RQ1 and RQ2. For instance, the *rule questionnaire* concentrates on the importance, value, comprehensibility, relevance and awareness of each rule, while the *assignment* and *completeness questionnaires* focus on the relationship of rules to principles, as well as on the complete coverage of design principles (identification of white spaces).

*B. Qualitative*

Due to the fact that the quantitative results depend to some extent on the domain and the specific technologies used (programming languages, frameworks, etc.), workshops will be held after the second or third iteration. The aim of such a workshop is to qualitatively discuss the role of rules and design principles, including the general acceptance of design principles as well as their operationalization with rules. Additionally, design enhancements within the source code will be investigated to gain awareness of the treatment of rule violations. From a methodical viewpoint, each workshop will rely on the data analysis method proposed in [9] that allows revealing answers to RQ3 in a structured manner.

*C. Threats to validity*

The research method of this work is to some extend based on thoughts and approaches from a similar experiment [5]. In [5], the authors discussed a problem when designing software-engineering experiments with a low number of participants. Termed *within-subject design*, the circumstance demands the use of all participants on all treatments to achieve satisfactory statistical power. Problems with *within-subject designs* include learning effects, the risk that the knowledge of participants increases between experiments. To reduce this risk and to prevent instrumentation threats to validity, materials used for the experiment must be comparable. Additionally, we plan to carry out our quantitative and qualitative analysis with different development teams from different organizations.

---

[2] http://www.sonarqube.org/

V. CURRENT STATE AND FUTURE WORK

As mentioned earlier in this paper, we have already conducted a survey to identify the relevance of design principles. We asked participants to rank design principles according to their importance and to list any missing ones. In the end, 104 participants from different engineering domains and with different job roles participated. Currently, this work is unpublished, but it will become part of our next publication. To briefly summarize, the five most important design principles according to this survey are the Single Responsibility Principle, Separation of Concerns, Don't Repeat Yourself Principle, Information Hiding Principle, and the Open-Closed Principle. Still highlighted as interesting are the Dependency Inversion Principle, KISS (Keep It Simple and Stupid) Principle, and YAGNI (You Ain't Gonna Need It) Principle.

MUSE has already proven valid for automatically measuring aspects of software design. For example, the tool was used to operationalize measures that are defined in an OMG (Object Management Group) quality standard [10]. Using the tool for various measurement tasks has shown that it is easy to configure and not limited to operationalizing design principles, which is the focus of this thesis. Current enhancements concentrate on enhancing the coverage of the design principles, on improving its stability, as well as on identifying and reducing false-positive rule violations. These enhancements are indispensable in order to control threats to validity that would arise when adapting the tool during its evaluation.

In addition, we are currently testing our above-described research method. In a software-engineering course at the Johannes Kepler University, three project teams of undergraduate students are developing a native Java App for Android OS, and they are required to deal with rule violations of five pre-selected design principles. What this means is that their source code is iteratively analyzed with MUSE. Afterwards, the findings of this analysis are presented to each team so that they can handle the violations in the next iteration. Each iteration ends, furthermore, with a survey including the questionnaires described in Section IV. This pilot-study helps us to identify shortcomings in the tool's implementation and to polish and enhance our questionnaires.

Starting in the middle of this year, we will conduct our first evaluation with an industrial partner. For this first step, we will focus on the design principles identified as important according to the design-principle survey. This first evaluation relies on using the fine-tuned version of MUSE and the experience gained from the student projects. After the first partner, two additional organizations are available and ready for collaboration.

VI. RELATED WORK

The idea to automate the assessment of object-oriented design is not new. For instance, [11] represents important work in this area, in which the authors defined a set of metrics for measuring object-oriented design. Based on this work, different authors built their own, metric-based approaches to detecting design problems, such as [12] [13]. However, single metrics are not the key to find meaningful design issues.

Because metrics have low entropy (information content) for understanding real design problems, a *detection strategy* for interpreting the results of measurement was proposed in [14]. This approach works on a higher level, namely the level of design principles, and is closer to real problems than would be single metrics. Additionally, a detection strategy leads to the direct identification of the real causes of quality flaws as reflected in flaws at the design level [14]. Nevertheless, at its core, the approach relies on a combination of metrics in the detection strategy, and it therefore also shares the previously discussed shortcomings of the metric-based approaches.

Besides detection strategies, a new approach called Method for Intensive Design Assessments (MIDAS) has been recently published [6]. MIDAS follows a three-view model for design problems. The first view, called the "ility"-based view, is based on six quality attributes. The second view is the view of design principles (in our meaning), which bridges the gap between violations of design principles and design problems. Finally, the third view is the constraint-based view, which allows the consideration of project-specific recommendations. The two major shortcomings of this approach are that (1) the quality model is fixed and is therefore inflexible and (2) the method relies on standard, static code-analysis tools, so the results can only be used as indicators for violations of design principles. It is attempted to compensate for the second shortcoming on the level of method, where design experts interpret the indicators and manually guide improvement actions.

In order to delimit this thesis research from related work, note that our concept of measuring design principles strives for completeness and not only to indicate design flaws. We are eager to fully cover design principles with rules, although we already know that some principles cannot be fully covered. Moreover, our design quality model gains higher flexibility from the application of the Quamoco meta-model for quality models [7]. The Quamoco meta-model lastly allows us to specify aggregation and evaluation formulae in order to directly compare the design quality of projects with each other. We have already gained experience with the evaluation of basic source-code quality [8].

## VII. CONCLUSION

The major planned contribution of this PhD is the systematic validation of the value of design principles in an industrial context as well as the applicability of automated measurement tools and design quality models for operationalizing the automatic evaluation of design principles. Besides the results of the validation, the PhD will provide an evaluation framework that can be used for other quality model related validation projects, as well as the completed and enhanced design quality model and the measurement tools.

ACKNOWLEDGEMENT

Initial (yet unpublished) work on the design quality model and the MUSE library was carried out by H. Gruber, C. Körner, A. Mayr, R. Plösch, and M. Saft.